\documentclass[%
preprint,
amsmath,amssymb,
aps,
prl,
floatfix,
]{revtex4-1}

\usepackage{amsmath}    
\usepackage{mathtools}
\usepackage{siunitx}

\DeclareMathOperator\erf{erf}
\usepackage{graphicx}   
\usepackage{verbatim}   
\usepackage{color}      
\usepackage{subfigure}  
\usepackage{hyperref}   
\begin{document}

\title{Interatomic potentials for ionic systems with density functional accuracy based on
charge densities obtained by a neural network}
\author{S. Alireza Ghasemi$^{1,*}$, Albert Hofstetter$^2$, Santanu Saha$^2$ and Stefan Goedecker$^2$}
\affiliation{$^1$ Institute for Advanced Studies in Basic Sciences, P.O. Box 45195-1159, Zanjan, Iran}
\affiliation{$^2$ Department of Physics, Universit\"{a}t Basel, Klingelbergstr. 82, 4056 Basel, Switzerland}

\email[]{aghasemi@iasbs.ac.ir}

\date{\today}

\begin{abstract}
Based on an analysis of the short range chemical environment of each
atom in a system, standard machine learning based approaches to the
construction of interatomic potentials aim at determining directly
the central quantity which is the total energy.
This prevents for instance an accurate description of the energetics
of systems where long range charge transfer is important as well as
of ionized systems.
We propose therefore not to target directly with
machine learning methods the total energy but an intermediate physical
quantity namely the charge density, which then in turn allows to
determine the total energy.
By allowing the electronic charge to distribute
itself in an optimal way over the system, we can describe not only
neutral but also ionized systems with unprecedented accuracy.
We demonstrate the power of our approach for both neutral and ionized NaCl clusters where
charge redistribution plays a decisive role for the energetics.
We are able to obtain chemical accuracy, i.e. errors of less than a
milli Hartree per atom compared to the reference density functional results.
The introduction of physically motivated quantities which are determined
by the short range atomic environment via a neural network leads also to
an increased stability of the machine learning process and transferability of the potential.
\end{abstract}

\maketitle

Atomistic simulations for materials are nowadays widely applied to understand and design materials.
A wide range of simulation methods exist, ranging from quasi exact many electron wavefunction methods~\cite{Martin2008}, over
density functional theory (DFT) calculations
\cite{Kohn1965} and semi-empirical quantum mechanical methods~\cite{Goringe1997,Papaconstantopoulos2003} to inter-atomic
potentials \cite{Tosi1964}.
For the different methods there are well known trade-offs between the
computational costs and the accuracy of the calculations.
In contrast to classical force fields,
density functional calculations give sufficient accuracy for a wide range of properties and are therefore the method of choice
in a huge number of studies.
Due to their computational cost, DFT simulations are however limited in practice to systems containing less than a few thousand atoms.
Since there is a great need to do highly accurate simulations for larger system there have been numerous efforts to
improve the accuracy of force fields, without increasing their cost too much.
It has been widely recognized that fixed charges limit the accuracy of the established standard force fields.
In the chemistry and biology community
polarizable force fields have therefore been developed~\cite{Dick1958}.
However such polarizable force fields allow only
for the displacement of charges, but do not allow for a true charge transfer over large distances.
Charge equilibration methods~\cite{Rappe1991,Wilmer2012} offer this possibility and have been
implemented in force fields such as ReaxFF~\cite{vanDuin2001}.

For covalent bulk materials such as carbon and silicon machine learning based
total energy schemes~\cite{Bartok2010,Behler2008} turned
out to give density functional accuracy at greatly reduced numerical cost.
For finite systems such as clusters, the construction of highly accurate machine learning based force fields is more difficult,
since atoms at a surface behave quite differently from atoms in the bulk.
It is not surprising that an analysis of the charge distribution of
NaCl clusters obtained from density functional calculations clearly reveals a charge transfer between atoms at the surface and in the
core of the cluster.
The fixed charges of $\pm 1$ electron used in the established force fields such as the Tosi-Fumi force
field~\cite{Tosi1964} are therefore clearly inadequate to describe such systems with high accuracy.
Since in the standard machine learning based interatomic potentials the energy of the whole system is
written as a sum of the energies of individual atoms~\cite{Behler2007}, and since the energy of an individual atom is
determined by the short range chemical environment, long range charge transfers as well as ionization can not well be described.

Constant charges are also present in the majority of force fields for bio-molecules and finding good charges for the use in such
force fields is highly non-trivial.
Even though the electrostatic part of the interaction energy is not the dominating one in these
force fields, it is to be expected that variable charges could also lead in this context to considerable improvements of the accuracy.
Environment dependent charges obtained from neural networks (NN)
were recently introduced~\cite{Artrith2011}, but this approach does not give the correct total charge 
of the system and can for instance lead to fluctuations of the total charge in molecular dynamics simulations.
The fact that the total charge can not be fixed also prevents the treatment of ionized systems. 
In addition, atomic reference charges must be provided in this approach. 
The extraction of such charges from ab initio calculations is always ambiguous to a certain extent.

A machine learning based approach, which in the spirit of our paper, does not directly aim at the total energy, but at an intermediate
physical quantity is the work of Snyder~\cite{Snyder2012,Snyder2013} and coworkers, where they propose to construct machine learning based
kinetic energy functionals.
Whereas in that work the charge density can in principle adopt any form we restrict our charge density
to be an approximate superposition of atomic charge densities.
In this way we exploit the well known fact that
the charge density in molecules, clusters and solids is given within a first approximation by a superposition of atomic charge densities.
The distribution of the electronic charge density is determined by atomic environment dependent electronegativities.
These electronegativities are in turn determined by the short range environment of the associated atom and can easily be
predicted by a neural network process.
To demonstrate the power of our approach we choose ionic clusters where a correct description of the charge density is essential
since bonding is mediated through charge transfer.
The fact that the charge distribution can redistribute itself over the whole cluster will allow us to treat both surface and
bulk atoms with very high accuracy and we
will demonstrate that density functional accuracy can be obtained with such an approach for clusters.
In contrast to conventional force fields, our approach also allows to describe
ionized systems without the need of any reparametrization.

\begin{figure}
\centering
\includegraphics[width=0.49\textwidth]{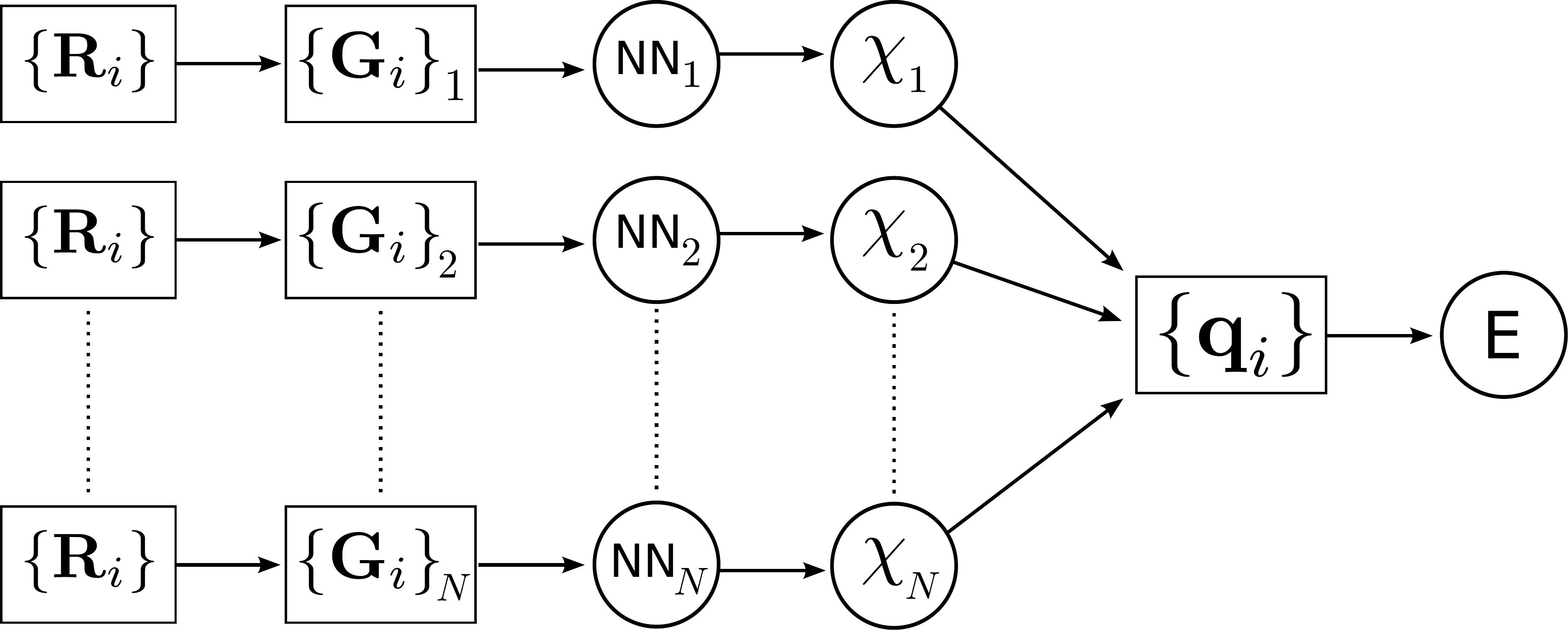}
\caption{Schematic illustration of the proposed method.
$\{{\bf R}_i\}$ and $\{{\bf G}_i\}$ are atomic positions and
symmetry functions, respectively.
\label{fig:schematic}}
\end{figure}

We consider a system consisting of $\frac{N}{2}$ Sodium (Na) and $\frac{N}{2}$ Chloride (Cl) atoms.
We postulate the following form for the total energy expression
\begin{align}\nonumber
U_{tot}(\left\{q_{i} \right\})=
&\sum_{i=1}^{N} \left(E_i^0 + \chi_{i} q_{i} + \frac{1}{2} J_{ii} q_i^2 \right) + \\ \label{eqn:etot}
&\frac{1}{2} \iint \frac{\rho({\bf r}) \rho({\bf r}')}{|{\bf r}-{\bf r}'|}
\;d{\bf r}\;d{\bf r}'
\end{align}
where $E_i^0$ are some fixed reference energies which in our implementation are
set to energies of isolated atoms, $q_i$'s are atomic charges,
$\chi_{i}$ is the  environment dependent atomic electronegativity of atom $i$
whose functional dependence is determined by a neural network approach.
To describe the hardness~\cite{Mortier1985} of atom $i$ we introduce element dependent atomic hardness terms
$J_{ii}$. 
The charge density of the system, $\rho({\bf r})$, is in our approach assumed to be a superposition of spherically symmetry Gaussian
functions centered at atomic positions, each normalized to the corresponding atomic charge $q_i$ given by
\begin{align*}
\rho_{i}({\bf r})=\frac{q_{i}}{\alpha_i^3 \pi^{\frac{3}{2}}} \exp\left(-\frac{|{\bf r}-{\bf r}_i|^{2}}{\alpha_i^2}\right).
\end{align*}
With this choice for the atomic charge densities, the total energy
of Eq.~(\ref{eqn:etot}) can be calculated analytically.
\begin{align}\nonumber
U_{tot}(\{q_{i}\},\{{\bf r}_{i}\})=
&\sum_{i=1}^{N} \left(E_i^0 + \chi_{i} q_{i} + \frac{1}{2}
(J_{ii}+\frac{2 \gamma_{ii}}{\sqrt{\pi}}) q_i^2 \right) + \\ \label{eqn:etotgauss}
&\sum_{i>j}^{N} q_{i} q_{j} \frac{\erf(\gamma_{ij} r_{ij})}{r_{ij}},
\end{align}
where $\gamma_{ij}=\frac{1}{\sqrt{\alpha_{i}^{2}+\alpha_{j}^{2}}}$ and $r_{ij}$ the distance between atoms $i$ and $j$.
The atomic charges $q_i$ are implicitly environment dependent through $\chi_{i}$ as will be explained below
and are therefore implicit functions of the atomic positions.
The implicit dependence of the $q_i$'s is obtained by minimizing the total energy of Eq.~(\ref{eqn:etotgauss})
with respect to the $q_i$'s.
This leads to a linear system of equations where all matrix elements depend on the atomic positions,
\begin{align}\label{eqn:linsys}
\frac{\partial U}{\partial q_{i}}=0,
\;\;\forall\; i=1,\ldots,N \;
\Longrightarrow
\sum_{j=1}^{N} A_{ij} q_j + \chi_i= 0.
\end{align}
where
\begin{align*}
A_{ij}=\left\{
\begin{array}{l}
J_{ii}+\frac{2 \gamma_{ii}}{\sqrt{\pi}}\;\;\;\;\text{for}\;\; i=j \\ \\
\frac{\erf(\gamma_{ij} r_{ij})}{r_{ij}}\;\;\;\; \text{otherwise.}
\end{array}\right.
\end{align*}
The fact that the  electrostatic interaction energy of any continuous charge density is always positive and that 
the $J_{ii}$'s are positive constants, implies that the matrix $A$ is positive definite and that the
linear system of equations has therefore always a unique solution which gives the minimal electrostatic energy.
The linear system Eq.~(\ref{eqn:linsys}) has to be solved under the constraint that
the atomic charges sum up to the correct overall charge $q_{tot}=\sum_{i=1}^{N} q_{i}$.
Adding this constraint via the Lagrange multipliers leads to the modified linear system of equations
\begin{align}\label{eqn:matrix}
\tilde{A} {\bf Q}={\bf X},
\end{align}
where ${\bf Q}$ and ${\bf X}$  are vectors  of the dimension $(N+1)$ and
$\tilde{A}$ is a $(N+1)\times(N+1)$ matrix.
In expanded form Eq.~(\ref{eqn:matrix}) is given as,
\begin{align}
\left( \begin{array}{ccc|c}
  &         &   & 1      \\
  & A_{i,j} &   & \vdots \\
  &         &   & 1      \\ \hline
1 &  ...    & 1 & 0      \\
\end{array}\right)
\left(\begin{array}{c}
q_{1}   \\
\vdots  \\
q_{N}   \\ \hline
\lambda \\
\end{array}\right)
=
\left(\begin{array}{c}
-\chi_{1} \\
\vdots    \\
-\chi_{N} \\ \hline
q_{tot}   \\
\end{array}\right)
\end{align}

\begin{figure}
\centering
\includegraphics[width=0.45\textwidth]{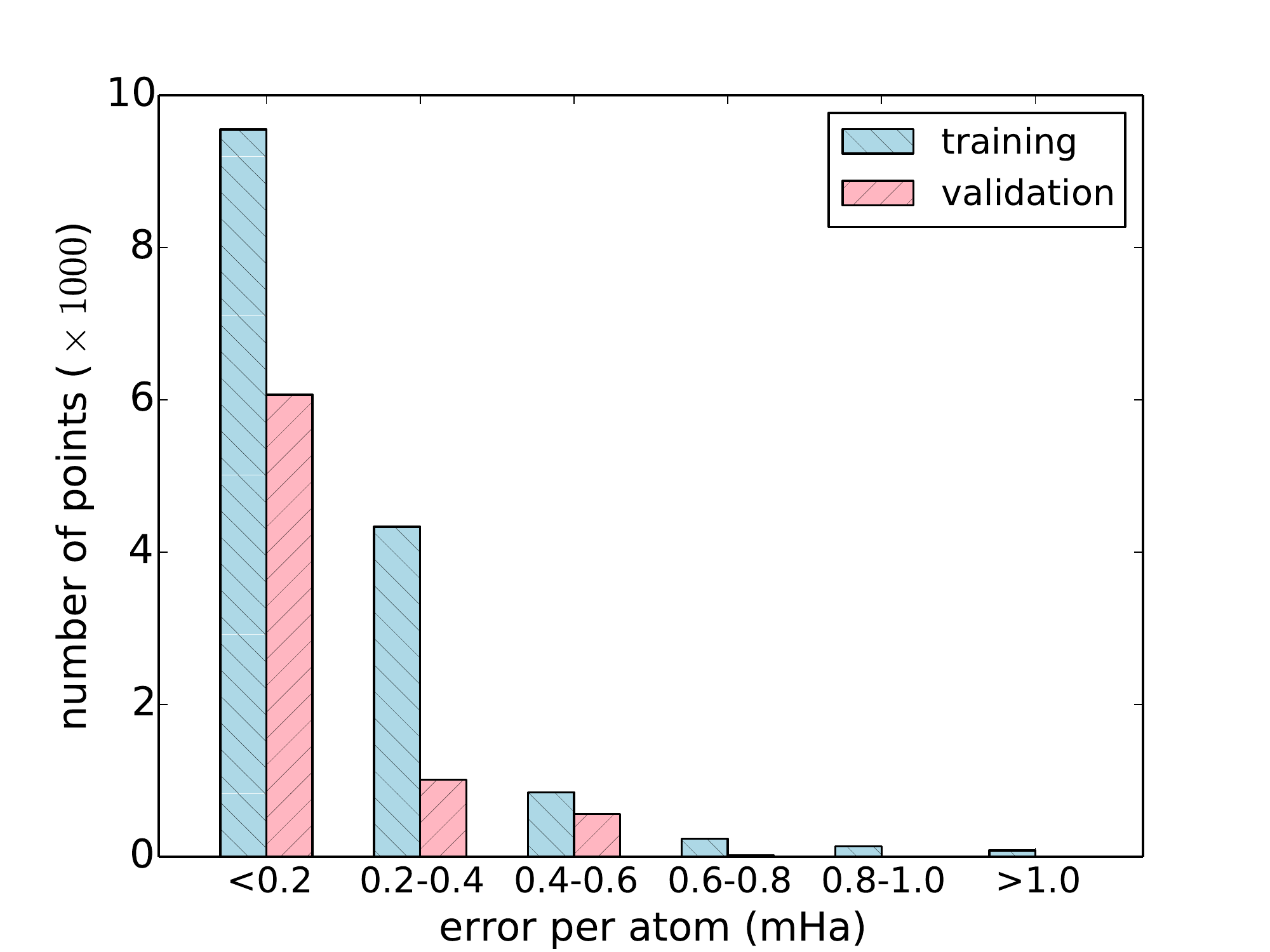}
\caption{(color online) Error distribution of the training and validation sets.
In total the training and validation sets contain $15191$ and $7658$ structures, respectively.
\label{fig:historgram}}
\end{figure}
In this way, we can allow for charge transfer over long distances without the need to find the 
presumably extremely complicated
explicit long range environment dependence of the $q_i$'s.
All that is needed to get the implicit long range  dependence of the $q_i$'s
on the atomic positions, is an explicit short range dependence of the $\chi_{i}$'s,
which is fixed once and for all by our neural
network together with the solution of a simple linear system of equations.
In addition, the total charge of the system is conserved unlike the method
given by Ref.~\cite{Artrith2011}.
This approach is physically motivated, since in a Kohn Sham density functional
calculation the total energy is minimized with
respect to the charge density distribution.
So the approach can be considered as some kind of constrained minimization of the total energy,
where the form of the charge density is restricted to be a sum over Gaussian functions with
amplitudes $q_i$.
In Kohn Sham density functional theory, the total energy consists of course not
only of the electrostatic term, but in addition there are the kinetic and
exchange correlation terms which oppose a charge distribution that would merely
minimize the electrostatic part of the total energy.
In our scheme this opposing force is represented by the constraints
on the form of the charge density and the environment
dependence of the $\chi_i$'s.

\begin{figure}
\centering
\includegraphics[width=0.45\textwidth]{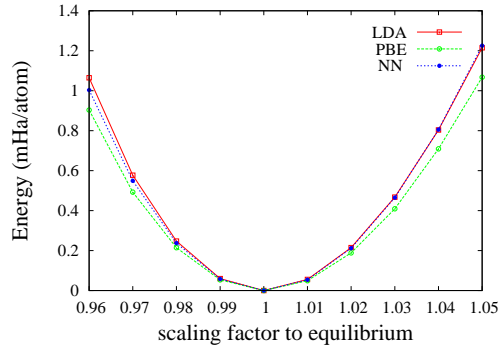}
\caption{(color online) Energy versus variation
of compression and expansion ratios relative to the equilibrium geometry of
the respective method for a $64$-atom NaCl cubic structure.
\label{fig:scale_NaCl064}}
\end{figure}

Fig.~\ref{fig:scale_NaCl064} shows LDA, PBE~\cite{Perdew1996} and NN energies
for different compression and expansion ratios relative to the equilibrium geometry of
the respective method for a $64$-atom NaCl cubic structure.
Even though our energy expression is missing the kinetic and exchange correlation terms,
the physically important energy differences can be calculated in exactly the
same was as in any density functional scheme based on the Hellmann-Feynman theorem. 
This theorem tells us that energy differences are given by the
integral over the atomic forces times the displacements.
The force acting on the atoms is the classical electrostatic force arising from
the charge distribution obtained through quantum mechanics.
Since our charge density is determined by a variational principle
it obeys the Hellmann-Feynman theorem.
In this way it is actually also granted that the charges obtained by
solving the linear system of equations give an accurate charge density within the
limitations imposed by our adopted form of the charge density.
Unreasonable charge densities would lead to wrong forces and hence
to wrong energy differences between different structures.

A schematic illustration of our method is given in Fig.~\ref{fig:schematic}.
Functional forms of symmetry functions ($\{G_i\}$) are those
given in Ref.~\cite{Behler2011} with a modification of the cutoff
function which is in our implementation a polynomial.
In order to construct and examine the method, we generated a large number of
reference data points using the BigDFT~\cite{Genovese2008} code.
An accurate evaluation of the electrostatic term
in density functional calculations is of great importance in particular for ionized clusters.
In the BigDFT code, the Hartree potential is calculated using the method
given in Ref.~\cite{Genovese2006} which enables us to have accurate
energetics for the ionized reference data points.
The DFT calculations are performed with the local density approximation (LDA).
A large number of data points consist of NaCl neutral and ionized ($+1$ and $+2$)
clusters ranging from $8$ to $80$
atoms with step sizes of $4$ are generated.
Data points are split into three sets; training, validation and test.
Clusters with fewer than $44$ atoms are considered in training set.
The structures of $64$-atom cluster are reserved for testing and the
rest are used as validation data set.

\begin{figure}
\centering
\includegraphics[width=0.45\textwidth]{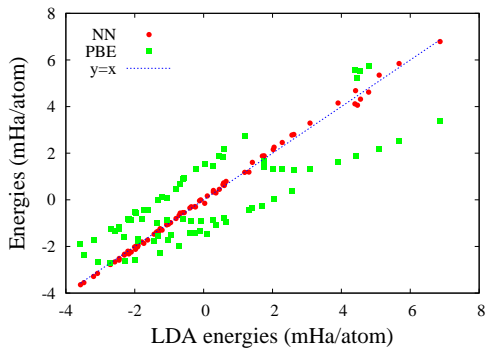}
\caption{(color online) Correlation of PBE/NN energies versus that of LDA
for neutral test points.
\label{fig:correlation}}
\end{figure}

In order to avoid overfitting, all training runs are performed with
only $15$ epochs.
The activation function of the output layer (values of electronegativities)
is taken to be the  $\tanh$ function rather than the commonly used linear
function.
This allows us to avoid a strong variation of atomic electronegativities
from one atom to another within a structure, which in turn may result in
too large variations in atomic charges.
In contrast to the standard NN potentials in Refs.~\cite{Behler2008,Artrith2011},
we can achieve in our approach small errors with a small number of nodes in NN hidden
layers.
Among several fits using various NN architectures and several
training runs with different initial random numbers for NN weights, we obtained the best compromise
between small RMSE and transferability of the potential for training and validation
data sets with the architecture $51\text{-}3\text{-}3\text{-}1$.
Gaussian widths of $1.0$ and $2.0$ bohr are consider for
sodium and chlorine atoms, respectively.
Values $0.2$ and $0.1$ in atomic units are found to be suitable for $J_{ii}$
for sodium and chlorine atoms, respectively.

The obtained RMSE of the training and validation sets is $0.26$ mHa per atom,
the RMSE of the neutral test set is $0.13$ mHa per atom and the RMSE of
the ionized test set (including $q_{tot}=+1$ and $q_{tot}=+2$) is $0.44$
mHa per atom.
The error distribution of the training and validation sets is illustrated in
Fig.~\ref{fig:historgram}.
For a few structures, the error is about $1$ mHa per atom while for
the majority of structures, the accuracy of our method is much higher than the so-called
chemical accuracy of 1kcal/mol.
Fig.~\ref{fig:scale_NaCl064} shows LDA, PBE~\cite{Perdew1996} and NN energies
for different compression and expansion ratios relative to the equilibrium geometry of
the respective method for a $64$-atom NaCl cubic structure.
The result shown in Fig.~\ref{fig:scale_NaCl064}
demonstrates clearly the transferability of our method since such compression and
expansion were neither included in the training nor in the validation set.

Fig.~\ref{fig:correlation} illustrates the correlation between PBE/NN and LDA energies.
All energies in Fig.~\ref{fig:correlation} are relative to the average energy
per atom of all structures for each method.
Correlation between NN and LDA energies is better than that between PBE and LDA
both for low- and high-energy structures, which shows that in principle even accuracies
higher than those obtainable from density functional theory could be achieved if the training/validation set energies
are calculated with a more accurate method.

In order to check whether the entire low energy configurational space is well described by our
interatomic potential we performed minima hopping~\cite{Goedecker2004} (MH) runs for different
cluster sizes.
MH explores in a systematic way this part of configurational space
and can thus detect whether the interatomic potential fails to describe
certain regions of this space.
No physically unreasonable configuration was found in all these runs indicating that
the interatomic potential describes well the entire low energy configurational space.

Based on a machine learning algorithm, a novel scheme is presented to reproduce with
high accuracy potential energy surfaces of quantum mechanical origin.
Instead of predicting atomic energies directly, we use NN methods to predict
environment dependent atomic electronegativities
from which atomic charges are obtained by a charge equilibration process~\cite{Rappe1991,Wilmer2012}.
Once the charges are available the total energy can be calculated easily.
Applying the method to neutral and ionized sodium chloride clusters
shows that unprecedented accuracy can be obtained for a particularly difficult system,
namely clusters where the atomic environment differs drastically between surface and
core atoms.
The error in energies of our interatomic potential compared to the reference
density functional data is far less than the error in total energies arising from
the use of different exchange correlation functionals.
The potential is highly transferable and
does not give rise to any unphysical structures.
\begin{acknowledgments}
Computational resources were provided by the CSCS under project s499 and the support of
Maria-Grazia Giuffreda is greatly acknowledged.
SS was supported by the SNF.
\end{acknowledgments}
\providecommand{\noopsort}[1]{}\providecommand{\singleletter}[1]{#1}%

\end{document}